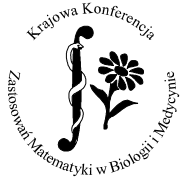

Leszno, 17th–20th September 2008

# CELLULAR AUTOMATA MODEL OF MACROEVOLUTION

Wojciech Borkowski[1]

[1]Center for Complex Systems in Institute for Social Studies,
University of Warsaw
ul. Stawki 5/7, 01-183 Warszawa
[1]borkowsk@samba.iss.uw.edu.

**ABSTRACT**

In this paper I describe a cellular automaton model of a multi-species ecosystem, suitable for the study of emergent properties of macroevolution. Unlike majority of ecological models, the number of coexisting species is not fixed. Starting from one common ancestor they appear by "mutations" of existent species, and then survive or extinct depending on the balance of local ecological interactions. Monte-Carlo numerical simulations show that this model is able to qualitatively reproduce phenomena that have been observed in other models and in nature.

## INTRODUCTION

As opposed to microevolution which works from the level of genes to the level of populations, the term macroevolution refers to biological evolution that occurs at, or especially, above the level of species. The so called Modern Synthesis School claims that this distinction is not important, and macroevolution could be understood as a longtime compound effect of microevolutionary processes. However, some theoretical biologists argue that evolution is a very *complex system* for which some macro processes or properties must be *emergent* and cannot be derived easily from the microevolution level (e.g. [1]).

The discussion remains open, among other things, because the macroevolution typically acts on a very long time scale, in the context of many ecosystems and the whole biosphere. Therefore, it is not amenable to experimental research – only post factum "historical" research is possible, and thus mathematical models and numerical simulations would be very important for the study of the macro level of evolution. They may be the only promising way to understand a variety of observable facts, repeatable and recurrent processes and patterns, and elucidate emergent consequences of such phenomena as Red Queen principle, key innovations, adaptive radiations, emergence or vanishing of natural barriers, climate change, local and global catastrophes, spontaneous or induced mass extinctions etc..

Unfortunately for "practicing" field biologists and paleontologists, the models currently dominating theoretical biology dismiss or omit different aspects of biological systems, which are obviously essential for the study of macro-evolution. Mathematical and computational ecology traditionally uses Lotka [2] Volterra [3] models that mostly disregard the spatial distributions of populations and always ignore their discrete character. This may lead to qualitatively improper predictions [4]. Additionally, such models can deal only with quite unrealistic systems composed of a few populations at most, the property, which becomes a serious drawback when they are applied to macroevolutionary processes.

The only widely known simulation model of macroevolution was proposed by Bak and Sneppen in 1993 [5] as a formal comment to the Gould's Punctuated Equilibrium hypothesis [6]. They showed how in a quite large set of interdependent species, self organized criticality might explain the main features of the whole fossil record (e.g. the distribution of sizes of extinction events). In the next few years the model was intensely disputed and is still explored by statistical physicists, bioinformatics and computer scientists (e.g. [7]). However, the Bak-Sneppen model necessarily oversimplifies the real mechanisms of speciation (the emergence of new species) and extinction events. The ecological adaptation and co-evolution of species are present there only in a very abstract sense and for this reason the model seems not to be applicable for more detailed macro-evolutionary questions.

Some ecologists and ecology oriented physicists and mathematicians recognize the limitations of the Lotka-Volterra based food (trophic) networks models. As an alternative, many kinds of micro-simulations were proposed (e.g. [8]), which were typically individual based and utilized a rectangular lattice as a substitute of environment, as reviewed by Pękalski [9]. In implementation they more or less resemble Cellular Automata or simple Agent Based Models that are also quite popular in social sciences. Most of these works investigate rather different aspects of predator-prey or predator-herbivore-plant systems (e.g. [10],[11]), but rarely they raise questions more related to macroevolution; like speciation (e.g. [12]) or the evolutionary caused avalanche of extinctions in multi predatory species ecosystems, where "over-specialized" predators evolve [13].

The simulation model presented in this paper belongs to this last group, however, with one important difference. Similarly to the famous Artificial Life simulation "Tierra" [14], but in a much simpler way the number of interacting species, both producers ("plants") and consumers ("herbivores", "predators" etc.), is not fixed and may accordingly increase or decrease during simulation. Every ecological niche possible at a particular moment of the simulation run might be taken by "speciation" event, occurring in existing populations. Such essential property gives this model a potential to deal with theoretical questions important for evolutionary biology, paleontology and even for astrobiology (e.g. when sources of biodiversity are deliberated).

## PRINCIPLES AND MODEL DEFINITION

My goal was to design an individual based model of macroevolution as similar as possible to a cellular automaton. I have utilized basic concepts, such as simple entities in partially occupied rectangular lattice[1], governed by local rules of interaction in Monte-Carlo dynamics, but the attributes of entities and the interaction rules are based on the central principles of Ecology, that are:

1. Understanding a *community* of species living in particular ecosystem as a network of energy (or biomass) flow - from producers (autotrophs) to a number of connected populations of consumers (heterotrophs);
2. The space of possible ecological niches is very large and multidimensional. It allows for many "ways of life" - both autotrophs and heterotrophs may have many adaptations for acquiring energy and for defending from exploitations. Because of energetic cost of those adaptations, organisms may also chose different levels of specialization (e.g. omnivores, herbivores, predators, annual plants, bushes, trees etc.)
3. The space of potential niches is searched by a process resembling the so called "random walk" – a newly emerged species takes a niche adjacent to the niche of its ancestral species. The local flow of energy plays a limiting role - each population in the ecosystem has to efficiently obtain enough energy from abiotic sources or from coexisting populations to at least balance the losses made by abiotic environment and by other populations; otherwise it would vanish later or sooner (rather).

---

[1] I use rectangles with side in proportion 2:1, not square like most of simulation designers do, because the effect of distance is often crucial for biological systems, and areas having the same size in two orthogonal dimensions are rather rare in nature.

4. Additionally, on a longer time scale, all populations are continuously evolving, pressed mainly by their enemies, competitors and the smartest prey (Red Queen principle) When, in changing conditions, a population fails in the balancing task, it becomes extinct, but such an event changes the environment for all interconnected species and causes reconstruction of the trophic network – sometimes leading to vanishing of other populations. I named the model CO-EVO to underline the essential role of multi-species co-evolution in long time evolutionary dynamics.

In practice, these principles are implemented in the model as follows:

Each site of the lattice could be empty or occupied by one simple entity, called "agent", which is characterized by 16-bit string defining its ability to interact with other agents, and two additional attributes: energy and age.

Each agent may be understood as one individual, belonging to one of possible classes, like in other CA models of ecosystems or, more abstractly, it may be treated as a small local subpopulation, which a solution allows a simpler interpretation of the model rules and results, especially because of the scaling problem.

In the course of simulation, the positions and states of agents change in Monte-Carlo dynamics, where agents are randomly selected for activity and the lattice sites subject to their actions are randomly chosen from the so called Moore neighborhood. Depending on the ecological characteristic and energy level of the active agent, and on the state of the selected adjacent site, whether it is empty or by whom it is occupied - a few actions are possible. The agent may move or set his offspring there, may just switch sites with the current owner or eat him, adding part of his energy to his own resources. Additionally, when an agent is selected, its age is incremented, and the agent dies after exceeding a particular limit (typically 50 Monte–Carlo steps). As a result, the agents that are able to survive, but are not able to reproduce, are prevented from an infinite persistence in the system.

Reproduction is formally asexual; therefore, in most of the cases, the progeny copies the ecological characteristic of the parent. Occasionally (on average 1 out of 10 or 1 out of 100 times for results referred below), the offspring would "mutate" by flipping one bit of its bitstring. Such technical solution is more or less similar to other evolutionary simulations and Genetic Algorithms; but, in absence of recombination, and because of an unusual interpretation of bitstrings (described below) and "populational" interpretation of agents such "mutations" events are also equivalent of tentative speciations. When a newly emerged "clone" is able to survive and proliferate locally in the lattice and becomes large enough, it is considered as a "species". This occurs when the clone exceeds 10 agents but this particular threshold has only technical meaning – just for filtering ecologically significant classes[2] which are undoubtedly able to survive for some time in their environment.

The main concept, which distinguishes this model from other evolutionary microsimulations (e.g. [10], [12], [13]), is the interpretation of the "genotype" bit-strings of an agent as whole ecological properties of it. This bit-string is divided into two 8-bit "masks". The first mask represents the ability to acquire resources; the second defines the ability to defend from exploitation. In general, both masks characterize the "overall ecological potential" of a particular agent because the bitwise similarity of the **acquiring mask** of active agent to the **defense mask** of potential prey determines the effectiveness of exploitation during interactions.

Particularly, the exploitation is possible when the acquiring mask of an active agent has at least one bit set at the same position as the defense mask of the attacked agent. The acquiring mask with all bits equal to 1 allows the agent to be an autotroph, but a 0 at any position causes the agent to be a heterotroph. An autotrophic agent acquires a number of energy units (50 for results presented below) every time when it is chosen by the M-C algorithm; whereas a heterotrophic agent must attack other agents with susceptible defense masks to gain energy. The prey always dies and only a fraction of its energy flows to the aggressor ($E_F$), whereas the rest of the energy disappears. The energy flow is calculated according to the formula below:

$$E_F = E_P \frac{M_P^d \cap M_A^a}{M_A^a} \frac{M_P^d \cap M_A^a}{M_P^d} \qquad (1)$$

---

[2] In fact, asexually reproducing organisms (like ,many of *Protista, bacteria* and even some plans) were identified by biologists in very similar manner – by correlation between morphology and ecological role.

Where $E_P$ is the energy reserve of the prey, $M_P^d$ is the defense mask of the prey, $M_A^a$ is the numerical value of acquiring mask of the aggressor and downcast semicircles represent a bitwise AND operation. Such a formula (1) assures that for each non-zero defense mask[3], a strictly specialized acquiring mask allows to take all energy, but more omnivorous aggressors never take full benefit from a prey somehow specialized in defense. For example, an omnivorous aggressor with $M^a$ **11101111** takes about 94% of energy from not armored prey with $M^d$ **11111111**, but 18% from armored prey **00101100** and only 5% or even 0.5% from heavily armored preys **10000000** or **00000001** respectively. But a predator specialized for a prey like the last one ($M^a$ **00000001**) is still able to take 100% of its energy. (see also Fig. 1).

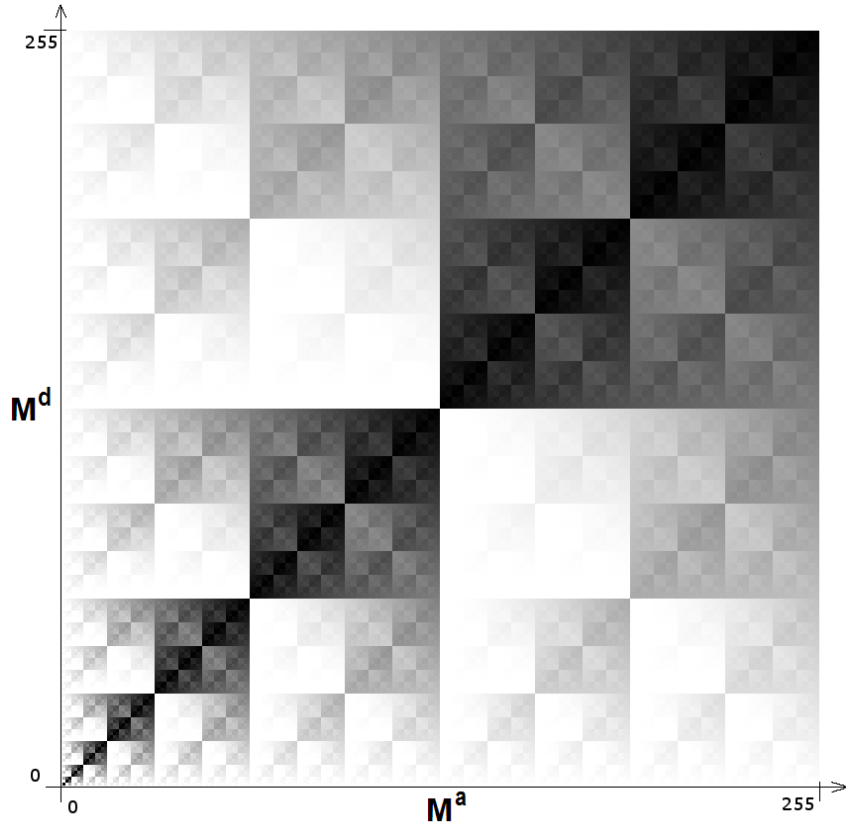

Figure 1. The space of exploitation intensity (*I*) - the fraction of energy which flows from a prey to an aggressor according to rule
$I = \frac{M^d \cap M^a}{M^a} \frac{M^d \cap M^a}{M^d}$, where $M^a$ is the aggression mask, $M^d$ is the defense mask of the prey. Each smallest rectangle represents in gray scale the value of *I* for interaction of particular aggression and defense masks, where white means very low flow of energy (0 or near 0), black means very high fraction (near 1 or 1).

---

[3] The defense mask equal to zero is forbidden, because it always causes an appearance of a completely immune autotrophic clone and a subsequent collapse of all others clones.

Some of the acquired energy is consumed every time, when an agent is chosen by M-C algorithm (1 unit for results in this paper). The remains are accumulated during the agent's life, and in suitable moments used to produce offspring. The parent equips each offspring with energy, by taking a fraction of their own reserve (typically 10%), but developing both the acquiring and the defense abilities come also at an energetic cost that is somewhat proportional to the "ecological potential" of the particular masks. It is calculated as a sum of a numerical value of the acquiring mask and a value of the bitwise negated defense mask. Thus, the most costly offspring is either an autotroph or omnivorous offspring having an acquiring mask with many positions switched on (e.g. autotrophic mask **11111111** costs 255 units), but also an offspring armored with a few-bits defense mask, especially when the most weighted bits are switched off (most expensive defense mask is **00000001** which costs 254 units).

Many CA and individual based simulations start from random states, even if researchers know that such a random state may be very far from the expected equilibrium. However, for evolutionary simulation, where many different equilibrium states are theoretically possible, such a common practice may lead to artificial results. In my model - more biologically realistic - each simulation course starts from one autotrophic agent which has no "defense adaptations". Subsequent mutations may destroy the autotrophic skills and produce a species that would take resources from others. At this moment, "evolutionary arms race" starts and a complete community of species may be formed. A particular history of the formation depends on the intricate interplay between simulation parameters (e.g. size of lattice, number of energy units per autotroph per M-C step, etc.) and random processes deciding when and where a particular species appears.

**MODEL BEHAVIOR**

The CO-EVO model, although based on different techniques and assumptions, may produce analogous results to other better known models.

On a short time scale, the simulation dynamics reproduce Lotka Volterra cycles, similar to the ones reported from individual based predator-prey models (as reported [9]; best e.g. in [10]). However, such a behavior is well visible only in extreme conditions when a strict domination of one autotrophic and one heterotrophic clone is observed. Wherefore, these cycles emerge rather at the beginning of the simulation course, especially when it is run on a very small lattice. Later in the simulation course or on larger lattices, one can also observe a pattern of expansion and extinction of competing clones typical for any evolutionary microsimulations (Ray's "Tierra" [14] or Adami's "Avida" [15] are good examples) and for some kinds of Genetic Algorithms.

On medium time scales we also observe evolutionary cycles corresponding to the results obtained by Lipowski [11]. A very efficient consumer (e.g. the most prevalent "herbivore" species) causes a fast local decrease in its autotrophic prey; hence its population also significantly decreases or even becomes extinct. The prey population rebuilds quite quickly and less efficient clones of consumers, immigrated or newly evolved, have ideal conditions to grow. Sooner or later an efficient clone immigrates from other regions of the lattice or reemerges by new mutations and closes such a cycle.

Moreover, the changes observable on a long time scale are comparable to the results of the Bak-Sneppen model [3]. In a broad range of parameters CO-EVO model repeats qualitatively the same dynamics (two examples on fig. 2), composed from few successive stages: exponential colonization of lattice, first dynamic equilibrium with the ecosystem based on relatively weakly armored autotrophs, "great extinction" caused by a series of "key innovations" in autotrophs defense mask and a second equilibrium with the ecosystem based on heavy armored autotrophs [16]. In any equilibrium state the system has a constant rate of species turn-over. However, during the exponential growth at the beginning of the simulation course and just after the avalanche mass extinction, new species are produced much faster, that resembles the predictions of "Punctuated Equilibrium" theory [4]. The extinctions and transformations observed here have a cause similar to the Bak-Sneppen avalanches which are, broadly speaking, "random disturbances" of the interconnected network of species. However, what is for Bak-Sneppen system almost completely random in fact, in the CO-EVO model comes from ecological interactions between species.

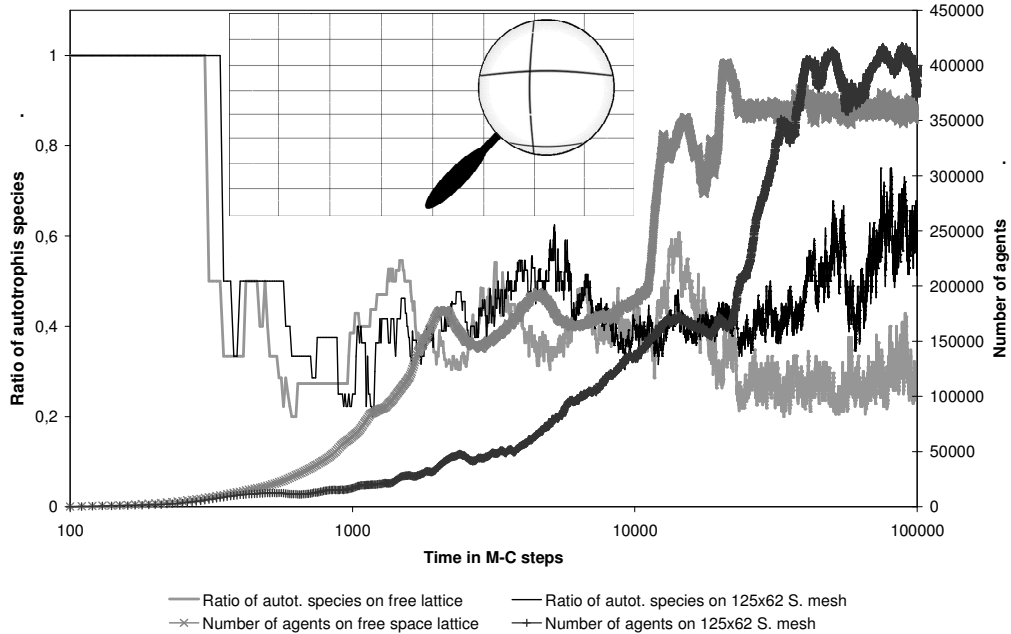

Figure 2. Histories of two simulations on 1000x500 lattices. Light gray lines represent a simulation performed on a lattice without any barriers, black lines represent simulation performed on very fragmentized lattice (as showed on the insert, by the 125x62 mesh with one cell width "communication pores" situated along an S-shaped curve, that restrict migration possibility but multiply the maximum possible distance). Time is in logarithmic scale to enhance visibility of the stages described in the text above.

Unlike most versions of the predator-prey and classical Bak-Sneppen models, my model does not artificially fix the number of "species" building the simulated ecosystem. It opens a possibility to examine relations between various global characteristics of an environment and an ecosystem complexity, simply measured by number of inhabiting agents and their classes (clones or species).

Many explorations of the model parameter space were done during model development[4] and previous work on it ([17],[16]), some more detailed study concentrated on robustness of the model dynamics, effects of the lattice size and autotrophs productivity will be published in a more elaborative paper [18]. Here, for reasons of limited space, I am able to present only one, however new, result. I examined how fragmentation of the environment introduced by rectangular mesh of finely pored barriers that changes the possibility of agents' dispersal, affects their number and proportion between autotrophs and heterotrophs.

---

[4] The model presented here is in fact quite old. Its first version was finished in 1995 and most of the core development was done in 1998 thanks to a "young scientist" grant of the Polish State Committee for Scientific Research.

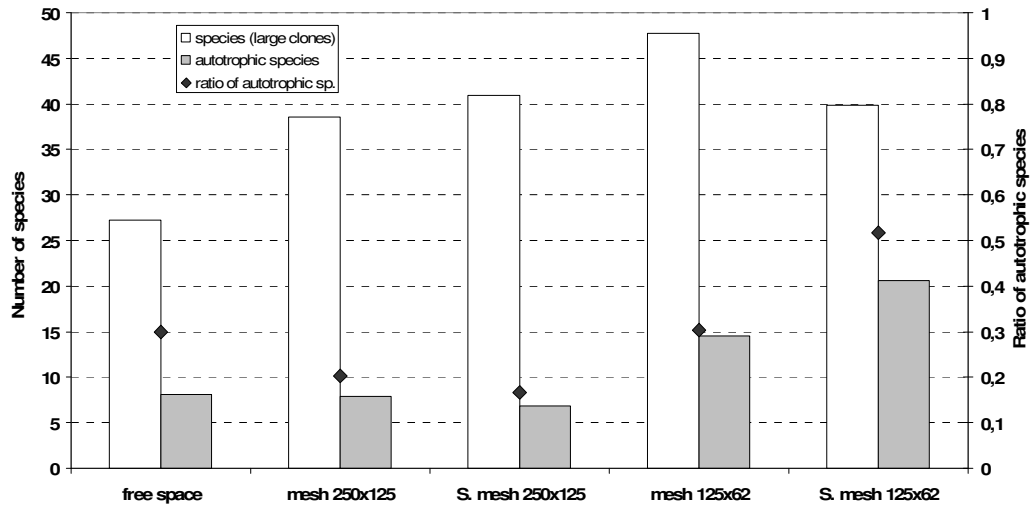

Figure 3. Mean number or species, autotrophic species and ratio of autotrophic species calculated from the second equilibrium of simulation courses performed on differently fragmented 1000x500 lattices. The fragmentation grows from left to right; **free space** means the lattice without any barriers; **mesh250x125** and **mesh125x62** mean fragmentation into patches sized 250 by 125 or 125 by 62 cells respectively, with "communication pores" (one cell width) in each side of the patch; **S.mesh250x125** and **S.mesh125x62** mean similar fragmentation, but with "pores" situated only in 2 sides of each patch, along an S-shaped curve, which restricts migration possibility and multiplies the maximum possible distance.

As one can see in Fig. 2, the whole dynamics of simulations significantly varying in such a parameter, does not differ qualitatively. The only definite differences are in durations of particular stages and the ratio of autotrophic agents during the second equilibrium. When we compare mean number or species, autotrophic species and ratio of autotrophic species taken from last one thousand M-C steps of the second equilibrium stage. (Fig. 3), it is clearly visible that fragmentation may increase the number of species, but within some range, and differently for autotrophic and heterotrophic species.

Such a result agrees less or more with basic biological knowledge - naturally fragmented ecosystems are usually richer than homogeneous ones, but additional anthropogenic fragmentation impoverishes them significantly; however, for further publishing in a biological journal, the fragmentation should be explored more finely, and results should be collated with real biological data.